\documentclass[a4paper,12pt]{article}

\usepackage{bbm}
\usepackage[utf8]{inputenc}
\usepackage{placeins}
\usepackage{amssymb,amsmath,amsfonts}
\usepackage[colorlinks,linkcolor=purple,citecolor=teal]{hyperref}
\usepackage{dsfont}
\usepackage{color}
\usepackage{graphicx}
\usepackage{hyphenat}
\usepackage{wrapfig}
\usepackage{empheq}
\usepackage{textcomp}
\usepackage[caption=false]{subfig}
\usepackage{rotating}
\usepackage{wrapfig}
\usepackage{pdfpages}
\usepackage{verbatim}
\usepackage{setspace}
\usepackage{array}
\usepackage{caption}
\usepackage{upgreek}
\usepackage{cmll}
\usepackage{latexsym}
\usepackage{braket}
\usepackage{cite}
\usepackage[left=2.3cm,top=3.2cm,right=2.3cm,bottom=3.2cm,bindingoffset=0cm]{geometry}
\usepackage[titletoc,page]{appendix}
\usepackage{multicol}
\usepackage{youngtab}
\usepackage{mathrsfs}
\usepackage{bm}
\usepackage[dvipsnames]{xcolor}
\usepackage{enumitem}
\usepackage{tensor}
\usepackage{slashed}
\usepackage{float}
\usepackage{tikz}
\usepackage{tikz-cd}
\usepackage{tcolorbox}

\newcommand{\af}{a_{\text{far}}}
\newcommand{\an}{a_{\text{near}}}
\newcommand{\Q}{\mathcal{Q}  }

\newcommand{\vf}{\varphi }
\newcommand{\rf}{ \text{ref} }

\newcommand{\beq}{\begin{eqnarray}}
\newcommand{\eeq}{\end{eqnarray}}
\newcommand{\bea}{\begin{eqnarray}}
\newcommand{\eea}{\end{eqnarray}}
\newcommand{\be}{\begin{equation}}
\newcommand{\ee}{\end{equation}}

\def\nn{\nonumber}

\allowdisplaybreaks
\tcbuselibrary{breakable}
\graphicspath{ {./Images/} }
\numberwithin{equation}{section}

\begin{document}

\title{
%\begin{flushright}\ \vskip -1.5cm {\small {IFUP-TH-202?}}\end{flushright}
\vskip 20pt
\bf{ On a soliton with \\ a non-normalisable zero mode}  
}
\vskip 40pt 
\vskip 40pt   
\author{
Guido Franchetti$^{1,2}$ and  Stefano Bolognesi$^{3,2}$ 
\\[15pt]	{\em \footnotesize
$^{(1)}$Department of Mathematics, University of Pisa}\\[-5pt]
{\em \footnotesize
Largo Pontecorvo, 5, 56127 Pisa, Italy}\\[2pt] 
{\em \footnotesize
$^{(2)}$INFN, Sezione di Pisa, 
Largo Pontecorvo, 3, Ed. C, 56127 Pisa, Italy}\\[2pt]
{\em \footnotesize
$^{(3)}$Department of Physics ``E. Fermi", University of Pisa}\\[-5pt]
{\em \footnotesize
Largo Pontecorvo, 3, Ed. C, 56127 Pisa, Italy}\\[2pt] 
{ \footnotesize   stefano.bolognesi@unipi.it,  guido.franchetti@unipi.it  }
} 
\date{}

\vskip 8pt
\maketitle

\begin{abstract}
We consider the holomorphic baby Skyrme model, which we propose as a simple tractable example of a soliton with a non-normalisable zero mode. The $1$-soliton has the same structure as a lump of fixed size. The size modulus of the sigma model becomes massive, while the phase modulus remains massless. The latter is non-normalisable in the $(0+1)$-dimensional sense because of the soliton tail. A suitable way to treat it is to consider the radial $(1+1)$-dimensional effective action on the half-plane. The soliton core sets the boundary condition for the effective model, which we evaluate with the WKB approximation and also numerically. We discuss the analogue of the effective moduli space approximation with a $1+1$ model and a boundary term at a certain cutoff scale.  This may serve as a prototype for more complex systems featuring non-normalisable zero modes such as certain kinds of monopoles in $3+1$ dimensions.
\end{abstract}

\newpage
\tableofcontents

\section{Introduction}

Soliton zero modes play a fundamental role in determining soliton dynamics. Under suitable conditions, namely low velocities and negligible radiation into massless fields, the dynamics is governed by an effective action for the zero modes. The metric on the moduli space is induced by their $L^2$ inner product. This approach has applications in various dimensions and to different types of solitons.

A static soliton in $d+1$ spacetime dimensions corresponds to a stable solution in $d$ Euclidean dimensions. If its zero modes are normalisable with respect to the $d$-dimensional $L^2$ norm, its low-energy dynamics can be reduced to a $(0+1)$-dimensional system \cite{Manton1982}. What happens when a zero mode is non-normalisable? The corresponding $(0+1)$-dimensional description is no longer useful: it assigns infinite inertia to the collective coordinate. Moving the entire soliton rigidly along the zero-mode direction would require infinite energy. A more physically meaningful question is what happens if we give the zero mode an initial push, with a velocity profile that decays sufficiently rapidly at large distances to have finite energy. Is there a low-energy effective description that is simpler than the full $(d+1)$-dimensional dynamics?

A systematic treatment and a common framework for non-normalisable zero modes are still lacking. The best-known example is provided by lumps in $(2+1)$-dimensional models with target space $S^2$ and a purely Dirichlet action. Another example is the magnetic monopole arising from the breaking $SU(2)\to U(1)$ in hyperbolic space, with spacetime $\mathbb{R}\times\mathbb{H}^3$, see for example \cite{Sutcliffe2022Boundary,FranchettiHarland2024}. A further important case is that of non-Abelian monopoles, for which the unbroken gauge group contains a non-Abelian subgroup, as in $SU(3)\to SU(2)\times U(1)$ or others, see for example \cite{Weinberg_2007,LeeWeinbergYi1996,ChenGuoWeinberg2001,Auzzi:2004if,Auzzi:2003em,Nitta:2010nd}. Different approaches to these models have been developed, motivated by both mathematical and physical considerations.

We first study what may be the simplest example of a soliton with a non-normalisable zero mode. In a particular baby Skyrme model, the soliton solution coincides exactly with a lump of the $S^2$ sigma model at a fixed size \cite{LeesePeyrardZakrzewski1990,Sutcliffe1991}. The size modulus is therefore stabilised, avoiding the difficulties it presents in lump dynamics \cite{Ward1985,Leese1990,Speight1995,SadunSpeight1998,SpeightStrachan1999}. The $S^1$ phase modulus remains, and its associated zero mode is non-normalisable.

We then consider $s$-wave fluctuations of the field around the soliton. We focus on the zero-angular-momentum modes associated with $S^1$ transformations. Away from the core, their radial equation has two linearly independent solutions that are  non-normalisable as $r\to\infty$. 
When continued into the soliton core, one solution is normalisable at $r=0$, while the other one is not. This leads to an effective $(1+1)$-dimensional theory for the radial $s$-wave $S^1$ mode, which can reach the core even at low energies.  This effective theory, together with a boundary condition set by the soliton core, plays the role of the moduli-space approximation for a non-normalisable mode. A spatially constant, rigid motion still has infinite energy, whereas the scattering modes of physical interest have finite energy.  Up to rescaling, the models form essentially a one-parameter family. We derive the boundary condition analytically in the low- and high-frequency regimes for arbitrary values of the parameter.

The paper is organised as follows. In Section~\ref{baby}, we review certain aspects of the holomorphic baby Skyrme model. In Section~\ref{effectivelag} we discuss the $S^1$ phase mode of this model.   We conclude in Section~\ref{conclusion}.

\section{From lumps to the holomorphic baby Skyrmion}
\label{baby}
The baby Skyrme model is a modified non-linear sigma model with additional terms added in order, among other things, to prevent solutions from shrinking to zero size in finite time.  We start by reviewing the pure non-linear sigma model mostly following \cite{manton:2004}. The non-linear sigma model on the plane with target the round sphere $S ^2$ has  Lagrangian 
\begin{equation}
\mathcal{L} =\frac{1}{4}  \partial _\mu \phi \cdot \partial ^\mu \phi 
\end{equation} 
where $\phi : \mathbb{R} ^2 \rightarrow S ^2 \subset \mathbb{R} ^3  $ must satisfy $| \phi |^2 =\phi _1 ^2 +\phi _2 ^2 +\phi _3 ^2 =1$. 
The associated static energy is
\begin{equation}
\label{energylump} 
U= \frac{1}{4} \int _{ \mathbb{R}  ^2 } \mathrm{d} \phi \wedge * \mathrm{d} \phi 
\end{equation} 
where $* $ is the Hodge operator on the domain $\mathbb{R} ^2 $.
We also  require $\phi$ to converge to a constant unit vector at spatial infinity along any direction,
\begin{equation} 
\label{phias} 
\phi _\infty =\lim _{ |x |\rightarrow \infty } \phi (x) .
\end{equation} 
Because of (\ref{phias})  we can replace the domain with its one-point compactification and view $\phi$ as a map $\phi : S ^2 \rightarrow S ^2 $. As such it has a Brouwer degree which for a smooth field can be calculated via
\begin{equation}
\label{degree2sphere} 
\operatorname{deg} \phi = \frac{1}{4 \pi } \int _{ \mathbb{R}  ^2} \phi ^\ast (\operatorname{vol} _{ S ^2 } ),
\end{equation} 
where
\begin{equation}
\operatorname{vol} _{ S ^2 } = \frac{1}{|x |^2 }\left( x _1 \mathrm{d} x _2 \wedge \mathrm{d} x _3 + x _2 \mathrm{d} x _3 \wedge \mathrm{d} x _1 + x _3 \mathrm{d} x _1 \wedge \mathrm{d} x _2 \right) .
\end{equation} 
Equivalently \begin{equation}
N := \operatorname{deg} \phi 
=\frac{1}{8 \pi } \int _{ \mathbb{R}  ^2 } \phi \cdot (\epsilon _{ ij }\partial _i \phi \times \partial _j \phi  ) \mathrm{d} x \wedge \mathrm{d}y,
\end{equation} 
where $\epsilon _{i j } =- \epsilon _{j i } $ and $\epsilon _{12 } =1 $. Since
\begin{equation*}
\begin{split} 
|\partial _i \phi \pm  \epsilon _{ ij } \phi \times \partial _j \phi |^2 &
=  2\partial _i \phi \cdot \partial _i \phi   \mp   2 \phi \cdot (\epsilon _{ ij }\partial _i \phi  \times \partial _j \phi ) ,
\end{split} 
\end{equation*} 
one obtains
\begin{equation}
0 \leq  \frac{1}{8} \int _{ \mathbb{R}  ^2 } |\partial _i \phi  \pm \epsilon _{ ij }  \phi \times \partial _j \phi| ^2 \mathrm{d} x \wedge \mathrm{d} y=
U\mp  2 \pi N .
\end{equation} 
Therefore
\begin{equation}
\label{bogbound} 
U\geq 2 \pi |N |,
\end{equation}
a Bogomolny-type bound, saturated if and only if 
\begin{equation}
\label{bogoforsigma} 
\partial _i \phi \pm \epsilon _{ ij }  \phi \times \partial _j \phi =0,
\end{equation}  
with the upper  (respectively lower) sign if $ N >0 $ ($N<0$). Solutions of (\ref{bogoforsigma}) are absolute minima of the energy in each given topological sector. Of course solutions of the first order equation (\ref{bogoforsigma}) also solve the second order variational equations associated to $U$. 
For $N =0 $ both equations (\ref{bogoforsigma})  need to be satisfied so  $\partial _i \phi =0 $ and $\phi=\phi _\infty $ is constant. This is the vacuum state of the theory.

It is  convenient to get rid of the constraint $|\phi |^2 =1 $ by using stereographic coordinates on the target space. Projecting from the south pole we define
\begin{equation}
\label{phistereo} 
w = \frac{\phi _1 + i \phi _2 }{1 + \phi _3 }.
\end{equation} 
We also write $ z =x + i y $ on the domain $\mathbb{R}  ^2 \simeq \mathbb{C}  $. 
Then
\begin{align}
\label{energystp} 
U&=\int _{ \mathbb{C}   } \frac{\mathrm{d}  w  \wedge * \mathrm{d} \bar w  }{(1 + | w  |^2)^2  } 
=i\int _{ \mathbb{C}  } \frac{ ( |\partial _z   w  |^2+ |\partial _{ \bar{z} }  w   |^2 ) \mathrm{d} z \wedge \mathrm{d} \bar{z}  }{(1 + | w  |^2)^2  }
\\ \label{topnumber} 
N &= \frac{i}{2 \pi }\int _{ \mathbb{C}   } \frac{ \mathrm{d}  w  \wedge \mathrm{d} \bar  w  }{(1 + | w  |^2 )^2 }
= \frac{i}{2 \pi }\int _{ \mathbb{C}  } \frac{ (|\partial _z  w  |^2 - |\partial _{ \bar{z} } w  |^2 ) \mathrm{d} z \wedge \mathrm{d} \bar{z}  }{(1 + | w  |^2 )^2 }.
\end{align} 
From (\ref{energystp}), (\ref{topnumber}) we reobtain $U \geq 2 \pi |N  |$, with equality if and only if $ w  $ is holomorphic  ($N>0$) or anti-holomorphic ($N<0$). These conditions can be checked to be equivalent to (\ref{bogoforsigma}).
For $N \geq 0 $ solutions of the Bogomolny equations, often called lumps, are thus meromorphic functions on $S ^2  $, that is rational functions
\begin{equation}
w (z) =\frac{ p (z)}{ q (z) }
\end{equation}
with $p \neq 0 $, $q$ coprime polynomials with complex coefficients. Without loss of generality we can take $\phi _\infty =(0,0,1) $, which translates to 
\begin{equation}
\label{basedrational} 
w (\infty ) =0  \Leftrightarrow  \operatorname{deg} (p)< \operatorname{deg} (q) ,
\end{equation} 
and $q$ to be a monic polynomial. The general solution  of degree $N$ then takes the form
\begin{equation}
\label{genlumphol} 
w (z )
= \frac{p _0 + p _1 z + \cdots + p _{ N -1 } z ^{ N -1 }}{q _0 + q _1 z + \cdots + q _{ N -1 }z ^{ N -1 } + z ^N }.
\end{equation} 
The $2 N $ complex parameters $\{ p _i , q _i \} $ are subject to the constraint imposing that the two polynomials have no common factor which is an open condition in $\mathbb{C}  ^{ 2 N }$. The moduli space of this sigma model is therefore a complex manifold of dimension $2  N $. Solutions with $N < 0 $ are obtained by taking the complex conjugate of (\ref{genlumphol}).

For $N =1 $ we have
\begin{equation}
\label{n1solpuresigma}
w (z)  =\frac{\lambda \mathrm{e} ^{i \chi } }{z-z _0 }
\end{equation} 
with $ z _0 \in \mathbb{C} $, $\lambda  \in (0, \infty )  $, $\chi \in [0, 2 \pi ) $. The energy of the single lump is
\begin{equation}
U= 2 \lambda ^2\int _{ \mathbb{C}  } \frac{ \mathrm{d} x \wedge \mathrm{d} y}{(|z-z _0  |^2 + \lambda   ^2 )^2 }
= 4 \pi \lambda ^2 \int _0 ^\infty \frac{\rho\, \mathrm{d} \rho }{(\rho ^2 +\lambda  ^2 )^2 } 
=2 \pi ,
\end{equation} 
having set $x - x _0  = \rho \cos \theta $, $y- y _0  =\rho \sin \theta $. The energy density $   (\rho ^2 + \lambda ^2 )^{ -2 }$ is strictly decreasing with $\rho$ with an absolute maximum at $\rho =0 \Leftrightarrow z = z _0  $ which can be interpreted as the position of the lump, giving two real moduli. The real parameter $\lambda $ can  be interpreted as the scale of the lump: exactly $\tfrac{1}{2} $ of the energy is contained in the disk of centre $z _0 $  and radius $ \lambda $. The fourth real modulus is the internal phase $\chi $.

Introducing time, the kinetic energy of the model is
\begin{equation}
T =\int _{ \mathbb{R} ^2   }\frac{|\dot w | ^2 }{(1 + |w |^2 )^2 } \mathrm{d} x \wedge \mathrm{d} y.
\end{equation} 
Consider the $N =1 $ solution (\ref{n1solpuresigma}). Following the adiabatic approximation, let us assume that $w$ depends on time only through its moduli.  If the only time dependent modulus is $\lambda$ then the kinetic energy is, using polar coordinates centred at $z _0 $,
\begin{equation}
T =2\pi  \dot \lambda ^2  \int _0   ^\infty  \frac{\rho ^3 \mathrm{d} \rho }{(\rho ^2 +  \lambda ^2 )^2 } =  2 \pi \dot \lambda ^2  \frac{1}{2} \left[ \log ( \lambda ^2 + \rho ^2 )- \frac{\rho ^2 }{ \lambda ^2 + \rho ^2 } \right] _0 ^\infty 
\end{equation} 
which is logarithmically divergent for large $\rho $. Similarly $T$ diverges if $\chi$ varies in time. The integral for $z _0 $ varying in time is finite instead. Hence in the adiabatic approximation the centre of the lump can move but its radius or phase cannot change.

For higher order solutions of the form $w =(\lambda / z ) ^N $  the kinetic energy is some constant multiple of $ \dot \lambda ^2 $ implying that, within the adiabatic approximation scheme, the parameter $\lambda$ can vary linearly in time, leading to lumps collapsing to a point in finite time \cite{LeesePeyrardZakrzewski1990Stability}. For $N=1$ the adiabatic approximation is not available since $\lambda$ has infinite inertia but collapse still occurs \cite{BizonChmajTabor2001,RaphaelRodnianski2012}. This behaviour can be traced to the invariance of the theory under conformal transformations of the domain. In order to stabilise the model  one needs to break this scale invariance.  In the baby Skyrme model \cite{PietteSchroersZakrzewski1995} the Lagrangian is modified by the addition of terms of order 4 and 0 in the derivatives,
\begin{equation}
\label{lagskyrme} 
\mathcal{L} =\frac{1}{4} \partial ^ \mu  \phi \cdot \partial _\mu  \phi - \frac{\eta }{4} (\partial _\mu \phi \times \partial _\nu \phi ) \cdot (\partial ^\mu \phi \times \partial ^\nu \phi )  - V (\phi ).
\end{equation} 
Here $\eta  $ is a positive constant and it is assumed that $|\phi |^2 =1 $. The topological structure of the baby Skyrme model is the same as that of the pure non-linear sigma model we just reviewed, but since all the additional energy contributions are, for a non-vacuum solution, positive, the Bogomolny bound (\ref{bogbound}) is never saturated.  The term of order 4 in the derivatives is known as the Skyrme term.  By Derrick’s argument, the Skyrme term alone does not allow for stable solitons in $2 + 1$ dimensions, hence the addition of the potential term $V (\phi) $.

The potential term is somewhat  arbitrary, with the qualitative behaviour of solutions depending strongly on it. Potentials which have been studied include
\begin{equation}
\label{genpotk} 
V (\phi) = C  \left( 1 -  n \cdot \phi \right)  ^k ,
\end{equation} 
with $C$ a positive constant and $n$ a unit vector. In this paper we will focus on the case $k =4 $,  known as the holomorphic model \cite{LeesePeyrardZakrzewski1990,Bolognesi:2014ova}. More precisely we will take
\begin{equation}
\label{genericpotential} 
V ( \phi )=  \frac{ \theta _4}{ 16 }  (1- \phi _3 )^4 .
\end{equation} 
Upon stereographic projection (\ref{phistereo})  the Lagrangian (\ref{lagskyrme}) becomes
\begin{equation} 
\label{lagrbabyskyrme} 
\begin{split} 
\mathcal{L}  &=
\frac{1}{ (1 + |w| ^2)^2   }  \left( | w _t |^2 - |w _x |^2 - |w _y |^2  \right)  \\ &
+  \frac{ 2 \eta }{(1 + |w| ^2 )^4 }   \left(   ( \bar w _x w _y  - w _x  \bar w _y  ) ^2 
-( \bar w _t w _x  - w _t  \bar w _x  ) ^2  - ( \bar w _t w _y  - w _t  \bar w _y  ) ^2\right)  \\ &
- \theta _4  \left( \frac{|w |^2 }{1 + |w |^2 } \right) ^4,
\end{split} 
\end{equation} 
where $w _t =\partial _t w $, etc.
The  equations of motion obtained from (\ref{lagrbabyskyrme})  are 
\begin{equation} 
\label{eom} 
\begin{split} 
w _{tt} - w _{xx} - w _{yy}
&- \frac{2 \overline{w }}{1 + |w |^2} \left( w _t^2 - w _x^2 - w _y^2 \right) \\ 
&+ \frac{4\eta }{(1 + |w |^2)^2}
\Big\{
2 \overline{w }_{ty} w _t w _y + 2 \overline{w }_{tx} w _t w _x - 2 \overline{w }_{xy} w _x w _y \\
&\qquad
+ \overline{w }_{xx}(w _y^2 - w _t^2)
+ \overline{w }_{yy}(w _x^2 - w _t^2)
- \overline{w }_{tt}(w _x^2 + w _y^2) \\
&\qquad
+ w _{xx}(|w _t|^2 - |w _y|^2)
+ w _{yy}(|w _t|^2 - |w _x|^2)
+ w _{tt}(|w _x|^2 + |w _y|^2) \\
&\qquad
+ w _{xy}(\overline{w }_x w _y + \overline{w }_y w _x)
- w _{tx}(\overline{w }_t w _x + \overline{w }_x w _t)
- w _{ty}(\overline{w }_t w _y + \overline{w }_y w _t)
\Big\} \\
&+ \frac{8\eta w }{(1 + |w |^2)^3}
\Big(
(\overline{w }_t w _y - \overline{w }_y w _t)^2
+ (\overline{w }_x w _t - \overline{w }_t w _x)^2
- (\overline{w }_x w _y - \overline{w }_y w _x)^2
\Big) \\
& +4\theta_4 \left( \frac{  |w |^2   }{1 + |w |^2} \right)^{3 }w
=0.
\end{split} 
\end{equation} 

A particular feature of the holomorphic model is that it retains a 3-parameter family of the solutions of the Bogomolny equations for the pure sigma model: the function
\begin{equation}
\label{nminus1lump}
w =\frac{\lambda \mathrm{e} ^{ i \chi } }{z- z _0  }
\end{equation} 
is a solution with $N = 1 $ provided that
\begin{equation}
\label{lambdac} 
\lambda  = \left( \frac{8 \eta  }{\theta _4 } \right) ^{ 1/4 }.
\end{equation} 
Hence compared with the $N =1 $ pure sigma model case, three of the four moduli survive while the lump's radius is fixed by (\ref{lambdac}). Just as for the pure sigma model, $\chi$ has infinite inertia. For $|N |> 1 $ rational functions do not solve (\ref{eom}). From now on we set $z _0 =0 $.

\section{Effective Lagrangian for the non-normalisable mode}
\label{effectivelag}  
Let us consider the ansatz
\begin{equation}
w =w _0  \mathrm{e} ^{ i \alpha (t, r ) }
\end{equation} 
where $w _0 $ is the unperturbed solution (\ref{nminus1lump}) and $\alpha$ a small perturbation. We are taking the perturbation to be radial only, i.e.~independent of the polar angle $\vartheta $ on the domain $\mathbb{R}  ^2 $. Substituting in (\ref{lagrbabyskyrme}) and dropping $ \alpha $-independent terms, which do not contribute to the equations of motion, gives
\begin{equation}
\mathcal{L}  =
\frac{\lambda^2 r ^2  }{(r^2 + \lambda^2)^2}
\left[\left( 1 + \frac{8 \eta  \lambda^2  }{(r^2 + \lambda^2)^2}\right) \alpha_t^2 - \alpha _r ^2  \right].
\end{equation}
Let us switch to the  dimensionless variable $x =r / \lambda $, $\tau = t/ \lambda $ and write $ h^\prime = \lambda \mathrm{d}h / \mathrm{d} r $, $ \dot h  = \mathrm{d}h / \mathrm{d} \tau  $.   Then 
\begin{equation}
\begin{split} 
\mathcal{L} \operatorname{vol} &
=\left[ \frac{x ^2 }{ (1 +x ^2 ) ^2 }  \left( \dot \alpha ^2 -  \alpha ^{ \prime 2 }  \right)  
+ \frac{\mu x ^2 }{ (1 +x ^2 ) ^4 }  \dot{\alpha} ^2   \right] x \,\mathrm{d} x \wedge \mathrm{d} \vartheta \\  &
= f (x) ( h (x) \dot\alpha ^2 -  \alpha  ^{ \prime 2} )\, \mathrm{d} x \wedge \mathrm{d} \vartheta ,
\end{split}
\label{actionalphafh} 
\end{equation} 
where
\begin{equation}
\mu = \frac{8 \eta}{\lambda ^2} 
\end{equation} 
and
\begin{equation}
\label{ourf} 
\begin{split}
f (x) &=\frac{x ^3  }{(1+ x^2)^2}, \quad 
h (x)= 1 +\frac{\mu }{(1 +x ^2 ) ^2 } .
\end {split} 
\end{equation} 
The corresponding equations of motion are
\begin{equation}
f h \ddot\alpha - (f \alpha ^\prime )^\prime 
= f (h \ddot \alpha - \alpha ^{ \prime \prime }) - f ^\prime \alpha ^\prime =0.
\end{equation} 
In particular if $\ddot \alpha =0 $ then
\begin{equation}
\alpha (x) = C _1  \int \frac{\mathrm{d} x}{f} + C _2 .
\end{equation} 
For $f$ in (\ref{ourf}) we get
\begin{equation}
\alpha (x) =C _1 \left( \frac{x ^2 }{2  } + 2 \log x - \frac{1 }{2 x ^2 } \right) + C _2 .
\end{equation} 
Stationary solutions of the form $\alpha =a (x) \exp (i \epsilon  \tau  )$, where $\epsilon = \lambda \omega $ is the dimensionless frequency, need to satisfy the ODE
\begin{equation}
\label{statodefull} 
\epsilon  ^2 f h a  +  (f a ^\prime )^\prime =0.
\end{equation}

\subsection{Studying the ODE}
We are now going to look at the approximate  solutions which can be obtained in the two limits $r \gg \lambda $ and $r \ll \lambda $, that is $x \gg 1 $ and  $x \ll 1 $. We first rewrite (\ref{statodefull}) in the form
\begin{equation}
\label{odek4bis} 
\frac{\mathrm{d}^2  a}{\mathrm{d} x ^2 } + P  (x)  \frac{\mathrm{d} a}{\mathrm{d} x} +\epsilon ^2 Q (x)  a =0,
\end{equation} 
with
\begin{equation}
\label{QPdef} 
\begin{split} 
P &=\frac{1}{f}\frac{\mathrm{d} f}{ \mathrm{d} x} = \frac{3- x ^2 }{x (1 + x ^2 )}, \qquad 
Q =1 + \frac{\mu }{(1 + x ^2 )^2 }.
\end{split} 
\end{equation} 

For  $x\ll 1$ we have
\begin{equation}
\mathcal{L}  \mathrm{vol} 
= \left[x ^3 \left(( 1 +  \mu )\dot\alpha^2  -\alpha^{\prime2} \right)  + O \left(x ^5  \right) \right] \mathrm{d} x \wedge \mathrm{d} \vartheta .
\end{equation} 
A stationary solution $\ddot \alpha =- \epsilon ^2 \alpha $ has to satisfy the approximate ODE 
\begin{equation}
a ^{ \prime \prime } + \frac{3 a ^\prime }{x} + \epsilon  ^2 (1 + \mu ) a =0.
\end{equation} 
The solution  can be written in terms of Bessel functions,
\begin{equation} 
\label{anear} 
\begin{split} 
\an  (x) &= \frac{1}{x} \Big[  C _1  J _1 (\epsilon ( 1 + \mu )^{ 1/2 } x ) + C _2 Y _1 (\epsilon ( 1 + \mu )^{ 1/2 }  x )  \Big].
\end{split} 
\end{equation} 
Since, for small $x$,
\begin{equation}
J _1 (K x )  = \frac{K }{2}  x  +O (x ^3 ) , \qquad Y _1 (K x ) =- \frac{2}{\pi K  \,x}  + \frac{Kx }{ \pi }  \log \left(\frac{Kx}{2} \right)  +\frac{Kx }{2 \pi }( 2 \gamma -1  ) +O (x ^3 \log x ),
\end{equation} 
where $\gamma$ is the Euler--Mascheroni constant, we need to take $C _2 =0 $ in order to have a solution which is well-defined at $x =0 $. Then $\an $ is constant to leading order near $x =0 $,
\begin{equation}
\an =   \frac{\epsilon  C _1 \sqrt{ 1 + \mu }}{2 }  + O (x ^2 ).
\end{equation} 
Since
\begin{equation}
\lim _{ x \rightarrow 0 } \left( x ^{-1}  J _1 (K x ) \right)  ^\prime =0,
\end{equation} 
while the derivative of $ x ^{-1} Y _1 (K x ) $ diverges as $x \rightarrow 0^+$, the condition $C _2 =0 $ is a boundary condition of Neumann type,
\begin{equation}
\label{neucond} 
\lim _{x \rightarrow 0 ^+} \an ^\prime (x) =0.
\end{equation}

In the opposite regime  $x\gg1$ we have
\begin{equation}
\mathcal{L} \operatorname{vol} 
= \left[\frac{1 }{x } (\dot{\alpha} ^2 - \alpha ^{ \prime 2 } )  + O  \left( x ^{-3 }  \right)\right] \mathrm{d} x \wedge \mathrm{d} \vartheta .
\label{effective1}
\end{equation} 
The corresponding stationary ODE is
\begin{equation}
\label{odefar} 
a ^{ \prime \prime } - \frac{a ^\prime }{x} + \epsilon ^2 a =0,
\end{equation} 
which has solution
\begin{align}
\label{afar} 
\af (x) &
= x \Big( A _1  J _1 (\epsilon  x )+ A _2   Y  _1  (\epsilon  x ) \Big) \\ &
\label{afarasy} 
=  \sqrt{  \frac{2x (A _1 ^2 + A _2 ^2 ) }{\pi \epsilon}}  \cos \left(  \epsilon x + \Theta    \right)  + O (x ^{ -1/2 }),
\end{align} 
where 
\begin{equation}
\label{phaseshift} 
\tan \Theta  =\frac{A _1 - A _2 }{A _1 +A _2 } \quad \Leftrightarrow \quad  \frac{ A _1}{ A _2 } =\frac{1 +\tan \Theta }{1- \tan \Theta }  .
\end{equation} 

Let us also look at the small $x$ behaviour of $\af $. To second order the Taylor series near $x =0 $ is
\begin{equation}
\label{afarnear} 
\af = - \frac{2 A _2 }{\pi \epsilon } +  \frac{ \epsilon  A _2}{ \pi }  x ^2 \log \left( \tfrac{x \epsilon }{2} \right)+ \frac{\epsilon }{2 \pi } \left[ \pi A _1  + ( 2 \gamma -1 ) A _2   \right] x ^2 
+ O (x ^4 ).
\end{equation} 
 In particular
\begin{equation}
\label{A2incond} 
\af (0) = - \frac{2 A _2 }{\pi \epsilon }.
\end{equation}

Neumann conditions (\ref{neucond})  for the ``near'' solution (\ref{anear})  imply a constraint for the ``far'' solution (\ref{afar}) and thus a relation between the two coefficients $A _1 $, $A _2 $. In Sections \ref{large-eps} and \ref{small-eps} we show how to derive an analytic approximation for this ratio in the large- and small-$ \epsilon $ regimes, see equations (\ref{ratiopred}) and (\ref{ratiosmallepsilon}). To check our analytical approximation we also calculate the ratio  numerically by integrating the exact  ODE (\ref{odek4bis}) using Mathematica. The initial conditions that we want to impose are $a ^\prime  (0) =0 $ and, say, $a (0) =1 $ (since the space of solutions is 1-dimensional, the value of $a (0) $  has no effect on the asymptotic ratio $A _1 / A _2 $).  
In practice, since the ODE coefficient $(\log f )^\prime $ diverges for $x \rightarrow 0  $, we impose the initial conditions at $x_0=10^{-4}$ using the expansion $\an(x)=1-\tfrac{1}{8}\epsilon^2(1+\mu)x^2+O(x^4)$, normalised so that $\an(0)=1$. We integrate (\ref{odek4bis})  up to $x_{\max}=x_{\rm fit}+12\pi/\epsilon$, where $x_{\rm fit}=\max(15,10/\epsilon)$, and determine $A_1$, $A_2$ by a least-squares fit of the solution to $x J_1(\epsilon x)$ and $x Y_1(\epsilon x)$ over the six periods in $[x_{\rm fit},x_{\max}]$. Note that for $\epsilon \ll 1 $  the lump core  $x \lesssim1 $ contributes a correction of order $\epsilon ^2 $ relative to $a (0) =1 $, so the integration tolerance must be well below this value.

\subsection{The large-$\epsilon$ regime}
\label{large-eps} 
For  $\epsilon\gg 1$ we can study the ODE (\ref{odek4bis}) using the WKB method. First write
\begin{equation}
\label{apsi} 
a = \psi \exp \left( - \frac{1}{2} \int P (x) \mathrm{d} x \right) =\left( \frac{1+ x ^2   }{x ^{ 3/2 }}\right) \psi  ,
\end{equation} 
under which the full ODE becomes
\begin{equation} 
\label{odewkbform} 
\psi ^{  \prime \prime } +  \Q \psi =0,
\end{equation} 
where
\begin{align}
\label{qformula}  
\Q&=  \epsilon ^2 Q -V , \\
V &= \frac{1}{2} P ^\prime + \frac{1}{4}  P ^2  =\frac{3}{4x ^2  } - \frac{8}{(1 + x ^2 )^2 }
\end{align} 
and $P $, $Q $ are given by (\ref{QPdef}).  For perturbed spherical Schrödinger operators of this type see  \cite{KostenkoTeschl2013}. We also define
\begin{equation}
n =\sqrt{ Q } =\sqrt{ 1 + \frac{\mu }{(1 + x ^2 )^2 } }, \qquad  n _0 =n (0) =\sqrt{ 1 + \mu }.
\end{equation}

Perturbative WKB expansion up to order $O (\epsilon ^{-1} ) $ for an ODE of the form (\ref{odewkbform})  gives
\begin{equation}
\psi = \frac{\sqrt{\epsilon } }{\Q ^{ 1/4 } } \left( K _1   \mathrm{e} ^{ i \phi } + K _2  \mathrm{e} ^{ - i \phi } \right) ,
\end{equation} 
where $K _1 $, $K _2 $ are arbitrary integration constants and 
\begin{align} 
\label{wkbphaseintegrand} 
\vf  &=    \epsilon  n + \frac{1}{\epsilon } \left( - \frac{V }{2n} +\frac{3 n ^{ \prime 2 }-2n n ^{ \prime \prime }}{8 n ^3 }  \right) + O (\epsilon ^{ -3 }),\\ 
\phi (x) &=\int    \vf (x)  \, \mathrm{d} x.
\end{align}

By (\ref{apsi}),  for large $x$, $\psi = \af/ \sqrt{x } $. We can then use (\ref{afarasy}) to write
\begin{equation}
\label{psiaswkb} 
\psi
= \sqrt{  \frac{2(A _1 ^2 + A _2 ^2 ) }{\pi \epsilon}} \, \cos   \left( \epsilon x + \Theta   \right)  + O (x ^{ -1 }),
\end{equation} 
hence the constant $\Theta$ can be read from the asymptotic phase shift of $\psi$,
\begin{equation} 
\Theta =\lim _{ x \rightarrow \infty } \phi (x) - \epsilon x.
\end{equation}
We want to calculate $\Theta $ more explicitly. Up to order $\epsilon ^0 $  the potential $\mathcal{Q} $ is strictly positive (no turning points), so the term $ \mathcal{Q} ^{1/ 4 } $ is real and does not contribute to the phase of $\psi$. We then simply have
\begin{equation}
\Theta = \epsilon I (\mu) ,
\end{equation} 
with
\begin{equation} 
I(\mu) =\int  _0  ^ \infty  (n-1 ) \mathrm{d}x=\int  _0  ^ \infty  \left( \sqrt{ 1 + \frac{\mu }{(1 + x ^2 )^2 }}  -1 \right)  \mathrm{d}x.
\end{equation} 
From (\ref{phaseshift})  we then see that $A _1 / A _2 $ is quasi-periodic in $\epsilon$ with (quasi) period 
\begin{equation}
T = \frac{ \pi }{| \mathrm{d}   \Theta   / \mathrm{d} \epsilon| }=\frac{\pi }{I (\mu) }.
\end{equation} 
In particular  $A _1 / A _2 $
has poles whenever $A _2 =0 \Leftrightarrow  \Theta  =\pi /4 \ (\operatorname{mod}  \pi )$. Due to the quasi-periodicity of $\Theta $ the poles are also quasi-periodic, with spacing given by $T$.

At  order $\epsilon ^{-1} $ the potential $\mathcal{Q} $  has a single zero $x _\ast $ given by
\begin{equation}
\label{turningpoint} 
x _\ast = \sqrt{ \frac{3}{4 (1 + \mu )} }\frac{1}{\epsilon } + O (\epsilon ^{ -3 }),
\end{equation} 
so there is a phase contribution from the tunneling region $0 \leq x \leq x _\ast $. For $x \geq x _\ast $ let us  define more generally the phase shift $\Theta (x) $  as
\begin{equation}
\label{Thetadef1}
\begin{split}  
\Theta (x) &= \Theta  (x _\ast ) +\int _{ x _\ast  }^x (\vf (t) - \epsilon )\mathrm{d} t \\ &
= \Theta (x _\ast ) +\int _{x _\ast } ^x  \left[\epsilon (n-1) + \frac{1}{\epsilon } \left( - \frac{V }{2n} +\frac{3 n ^{ \prime 2 }-2n n ^{ \prime \prime }}{8 n ^3 }  \right)  \right] \mathrm{d} t+ O (\epsilon ^{ -3 }),
\end{split} 
\end{equation} 
where  $\Theta (x _\ast )$ represents  the contribution to the asymptotic phase due to the tunneling region $0 \leq x <  x_\ast $. Note that $\Theta \equiv \Theta (\infty )$.  

Near $x =0 $ we have
\begin{equation}
\label{potentialorigin}
\Q = \epsilon^2(1+\mu) + 8 - \frac{3}{4x^2} + O(x^2),
\end{equation}
so the centrifugal term $-\tfrac{3}{4x^2}$ dominates  and cannot be treated perturbatively. Following a Bessel comparison-equation method \cite{Olver1997} of a centrifugal singularity, we approximate $\psi$ near $x=0$  by the exact solution of a reference ODE carrying an identical singularity. We therefore introduce 
\begin{gather}
\psi_\rf^{\prime\prime} + \Q_\rf\,\psi_\rf = 0,\\
\label{qref}
\Q_\rf = \kappa^2 - \frac{3}{4x^2},\\
\label{xkappa}
\kappa = \sqrt{\epsilon^2(1+\mu)+8} = n_0\epsilon + \frac{4}{n_0\epsilon} + O(\epsilon^{-3}),
\end{gather}
whose potential reproduces the leading $x\to0$ behaviour of $\Q$ in \eqref{potentialorigin}.
The solution of the ODE
\begin{equation}
\psi^{\prime\prime} + \left(\kappa^2  - \frac{3}{4 x ^2 }\right) \psi   = 0
\end{equation}
which is  regular at $x =0 $ is
\begin{equation}
\label{asybesselj}
\sqrt{x}\,J_1(\kappa x) \simeq \sqrt{\frac{2}{\kappa\pi}}\cos\!\left(\kappa x - \frac{3\pi}{4}+ \frac{3}{8 \kappa x}\right).
\end{equation}
The potential $\Q_\rf $ has, to this order, the same turning point (\ref{turningpoint}).
Let $x _m  $ be a point such that $x _m \ll \epsilon ^{-1/3 }  $, so that using the solution $ \sqrt{ x }J _1 (\kappa x ) $ of the approximate ODE for small distances is justified, and $\epsilon x_m  \gg1 $, so that using the large distance asymptotics (\ref{asybesselj}) is justified. Then we can write 
\begin{equation} 
\label{phaseasysmall} 
\phi (x _m )=- \frac{3\pi }{4} + \kappa  x _m   + \frac{3}{8 \kappa  x_m } + o (1 )
= - \frac{3\pi }{4} + \epsilon n _0   x _m   + \frac{3}{8    x_m \epsilon n _0  }  + \frac{4x _m }{\epsilon n _0 }  + O (\epsilon ^{ -3 }),
\end{equation} 
where the first equality comes from identifying $\phi (x _m ) $ with the argument of $\cos $ in (\ref{asybesselj}), and the second one uses (\ref{xkappa}).
In particular the WKB phase $\phi_\rf $  accumulated by the reference solutions on top of its asymptotic   $\kappa x $ oscillatory behaviour is
\begin{equation}
\lim_{x\to\infty}\big(\phi_\rf(x) - \kappa x\big) = -\frac{3\pi}{4}.
\end{equation}
Supplying this constant, which is the contribution of the tunneling region to the WKB phase, is the reason for introducing $ \psi _\rf $.   We then calculate
\begin{equation}
\begin{split}
\label{thetapieces} 
\Theta &=
\Theta  (\infty )- \Theta (x _m )+ \Theta (x _m ) 
= \Theta (x _m ) +\int _{ x _m }^\infty \left(\vf -\epsilon \right) \mathrm{d} x .
\end{split}
\end{equation} 
The term $\Theta (x _m )$ represents the phase accumulated up to $x _m $. Since $\epsilon ^{-1} \ll x _m \ll \epsilon ^{-1/3 } $, its value can be read from the asymptotic behaviour (\ref{phaseasysmall}) of the small $x$ solution: 
\begin{equation} 
\Theta (x _m ) =\phi  (x _m ) -\epsilon  x _m =- \frac{3\pi }{4} + \epsilon  n _0  x _m   + \frac{3}{8 \epsilon n _0   x_m } + \frac{4x _m }{\epsilon n _0 } - \epsilon x _m  + o (1 ).
\end{equation}

We now calculate the integral in (\ref{thetapieces}).   By (\ref{wkbphaseintegrand}), it is given by
\begin{equation}
\int _{ x _m }^\infty (\vf - \epsilon )\mathrm{d} x 
=\int _{ x_m }^\infty  \left[  \epsilon(n-1) - \frac{1}{2 \epsilon n} V + \frac{1}{8 \epsilon } \left( \frac{3 n ^{ \prime 2 }}{ n ^3 } - \frac{2 n ^{ \prime \prime }}{n ^2 }\right) \right] \mathrm{d} x + O (\epsilon ^{ -3 }).
\end{equation} 
The first term is
\begin{equation}
\epsilon\int _{ x _m }^\infty ( n -1 )\mathrm{d} x
=\epsilon\int _{ 0 }^\infty (   n -1 )\mathrm{d} x - \epsilon\int _{ 0 }^{x _m } ( n -1 )\mathrm{d} x
=\epsilon I  (\mu)  - \epsilon (n _0-1)x _m   + O( \epsilon x _m ^3 )
\end{equation} 
since in the second integral $x \leq x _m $ and
\begin{equation}
n =\left( 1 + \frac{\mu }{(1 + x ^2 )^2 } \right) ^{ 1/2 }=  \left( 1 + \mu (1 - 2 x ^2 )+ O (x ^4 ) \right) ^{ 1/2 }
= n _0  + O (x ^2 ).
\end{equation} 
The $V$ term gives
\begin{equation}
\begin{split}
& - \frac{1}{2 \epsilon } \int _{ x _m } ^\infty  \frac{V }{n} \mathrm{d} x
=- \frac{1}{2 \epsilon } \int _{ x _m } ^\infty \left( \frac{3}{4n x ^2 }-\frac{8}{n(1 + x ^2 )^2 } \right) \mathrm{d} x\\ &
=- \frac{1}{2 \epsilon } \int _{ x _m } ^\infty \left( \frac{3}{4 x ^2 } \left( \frac{1}{n} - \frac{1}{n _0 } \right) -\frac{8}{n(1 + x ^2 )^2 } \right) \mathrm{d} x - \frac{3}{8 n _0 \epsilon x _m  }\\ &
=- \frac{1}{2 \epsilon } \int _{ 0} ^\infty \left( \frac{3}{4 x ^2 } \left( \frac{1}{n} - \frac{1}{n _0 } \right) -\frac{8}{n(1 + x ^2 )^2 } \right) \mathrm{d} x
+ \frac{1}{2 \epsilon } \int _{ 0} ^{x _m}  \left( \frac{3}{4 x ^2 } \left( \frac{1}{n} - \frac{1}{n _0 } \right) -\frac{8}{n(1 + x ^2 )^2 } \right) \mathrm{d} x
- \frac{3}{8 n _0 \epsilon x _m  }\\ &
=- \frac{1}{2 \epsilon } \int _{ 0} ^\infty \left( \frac{3}{4 x ^2 } \left( \frac{1}{n} - \frac{1}{n _0 } \right) -\frac{8}{n(1 + x ^2 )^2 } \right) \mathrm{d} x
+ \frac{3 \mu x _m }{8 \epsilon n _0 ^3 } - \frac{4 x _m }{ \epsilon n _0 }  - \frac{3}{8 n _0 \epsilon x _m  }+ O ( x _m ^3/ \epsilon  ).
\end{split}
\end{equation} 
Finally we have the $O (\epsilon ^{-1} )  $  WKB contribution,
\begin{equation} 
\begin{split} 
&\frac{1}{8 \epsilon } \int _{ x _m }^\infty \left( \frac{3 n ^{ \prime 2 }}{ n ^3 } - \frac{2 n ^{ \prime \prime }}{n ^2 }\right) \mathrm{d} x
=\frac{1}{8 \epsilon } \int _{ 0}^\infty \left( \frac{3 n ^{ \prime 2 }}{ n ^3 } - \frac{2 n ^{ \prime \prime }}{n ^2 }\right) \mathrm{d} x
-\frac{1}{8 \epsilon } \int _0 ^{ x _m } \left( \frac{3 n ^{ \prime 2 }}{ n ^3 } - \frac{2 n ^{ \prime \prime }}{n ^2 }\right) \mathrm{d} x\\ &
=\frac{1}{8 \epsilon } \int _{ 0}^\infty \left(  - \frac{2}{n ^{ 1/2 }} \mathrm{d}  \left( \frac{n^\prime }{ n ^{ 3/2 }} \right) \right) -\frac{ \mu }{2 \epsilon  n _0 ^{ 3 }} x _m  +  O(x _m ^3 )
=- \frac{1}{8 \epsilon } \int _0 ^\infty  \frac{n ^{ \prime 2 }}{n ^3 } \mathrm{d} x
-\frac{ \mu }{2 \epsilon  n _0 ^{ 3 }} x _m  +  O(x _m ^3 / \epsilon ).
\end{split} 
\end{equation} 
Summing the three terms gives
\begin{equation}
\begin{split}
\Theta (\infty )&
= - \frac{3\pi }{4}  + \epsilon I(\mu)
- \frac{1}{2 \epsilon } \int _{ 0} ^\infty \left( \frac{3}{4 x ^2 } \left( \frac{1}{n} - \frac{1}{n _0 } \right) -\frac{8}{n(1 + x ^2 )^2 } \right) \mathrm{d} x
- \frac{1}{8 \epsilon } \int _0 ^\infty  \frac{n ^{ \prime 2 }}{n ^3 } \mathrm{d} x\\ &
- \frac{ \mu x _m }{8 \epsilon n _0 ^3 }.
\end{split}
\end{equation} 
The last term is actually $o (\epsilon ^{ -1   })$, so we can drop it. In fact in the intermediate region  $\epsilon ^{-1} \ll x _m  \ll1 $ hence $x _m / \epsilon \ll 1/\epsilon \Rightarrow x _m / \epsilon = o (\epsilon ^{-1} )$ (the $\epsilon ^{-1} \sim x _\ast \ll x _m $ inequality is not needed here but justifies using the oscillatory asymptotics of Bessel’s function).

Hence defining
\begin{equation}
\begin{split}
\mathcal{I} (x)  &=\int  _0 ^x  (n-1 )\mathrm{d}t , \quad I =\mathcal{I} (\infty )\\
\mathcal{K}_1 (x)    & 
=-  \int _{ 0 } ^x \left( \frac{3}{8 t ^2 } \left( \frac{1}{n} - \frac{1}{n _0 } \right) -\frac{4}{n(1 + t^2 )^2 } \right) \mathrm{d} t, \quad K _1 = \mathcal{K}  _1 (\infty )\\
\mathcal{K} _2(x)  &
=- \frac{1}{8  } \int _0 ^x  \frac{n ^{ \prime 2 }}{n ^3 } \mathrm{d} x 
= - \frac{\mu ^2 }{2} \int _0 ^x  \frac{t ^2 \mathrm{d} t}{(1 +t^2 )^6  n ^5 }, \quad K _2 =\mathcal{K} _2 (\infty ),
\end{split} 
\end{equation} 
we have
\begin{equation}
\label{Thetaeq} 
\Theta =- \frac{3 \pi }{4} +    \epsilon I  + \frac{1}{\epsilon } (K _1 + K _2 ) + o(\epsilon ^{-1} ).
\end{equation}
When convenient we will consider $I $, $K _1 $, $K _2 $ as functions of the parameter $\mu$.

It is worth noting that this treatment shows that, to the order we are working at, we can define
\begin{equation} 
\Theta (x _\ast )= - \frac{3\pi }{4} + \epsilon \mathcal{I} (x _\ast )  + \frac{1}{\epsilon  } (\mathcal{K} _1 (x_\ast) + \mathcal{K} _2 (x_\ast) )
\end{equation} 
and consequently
\begin{equation}
\Theta  (x) = - \frac{3 \pi }{4} + \epsilon \mathcal{I} (x)  + \frac{1}{\epsilon  } (\mathcal{K} _1 (x) + \mathcal{K} _2 (x) ).
\end{equation} 
We have thus managed to express the phase contributions from both the tunneling and oscillatory regions as the integral of the same smooth function.  Interestingly we  then see that
\begin{equation}
\Theta  (0) =- \frac{3 \pi }{4},
\end{equation} 
so  the regularity choice  $ C _2 =0 $ at $x =0 $, see (\ref{asybesselj}), can be interpreted as setting the initial value of $\Theta$. The equation
\begin{equation} 
\label{ratiopred} 
\frac{ A _2 }{A _1} = - \tan  \left( \Theta  + \frac{ 3 \pi }{4 } \right)  = - \tan \left( \Theta  (\infty ) - \Theta (0)  \right)  
\end{equation} 
shows how $A _1 / A _2 $ can be viewed as a phase shift.

The following Figures \ref{wkbratio}, \ref{wkbphase} show the ratio $A _1 / A _2 $ and phase $ \cot\Theta $ obtained from the numerics and from the WKB estimate. The plots are shown for the value $\mu =1 $. The values of the three integrals are
\begin{equation}
I  (1)   \simeq 0.347, \qquad 
K _1 (1) \simeq  2.300, \qquad 
K _2 (1) \simeq -0.007.
\end{equation} 

\begin{figure}[htbp]
\begin{center}
\includegraphics[scale=0.4]{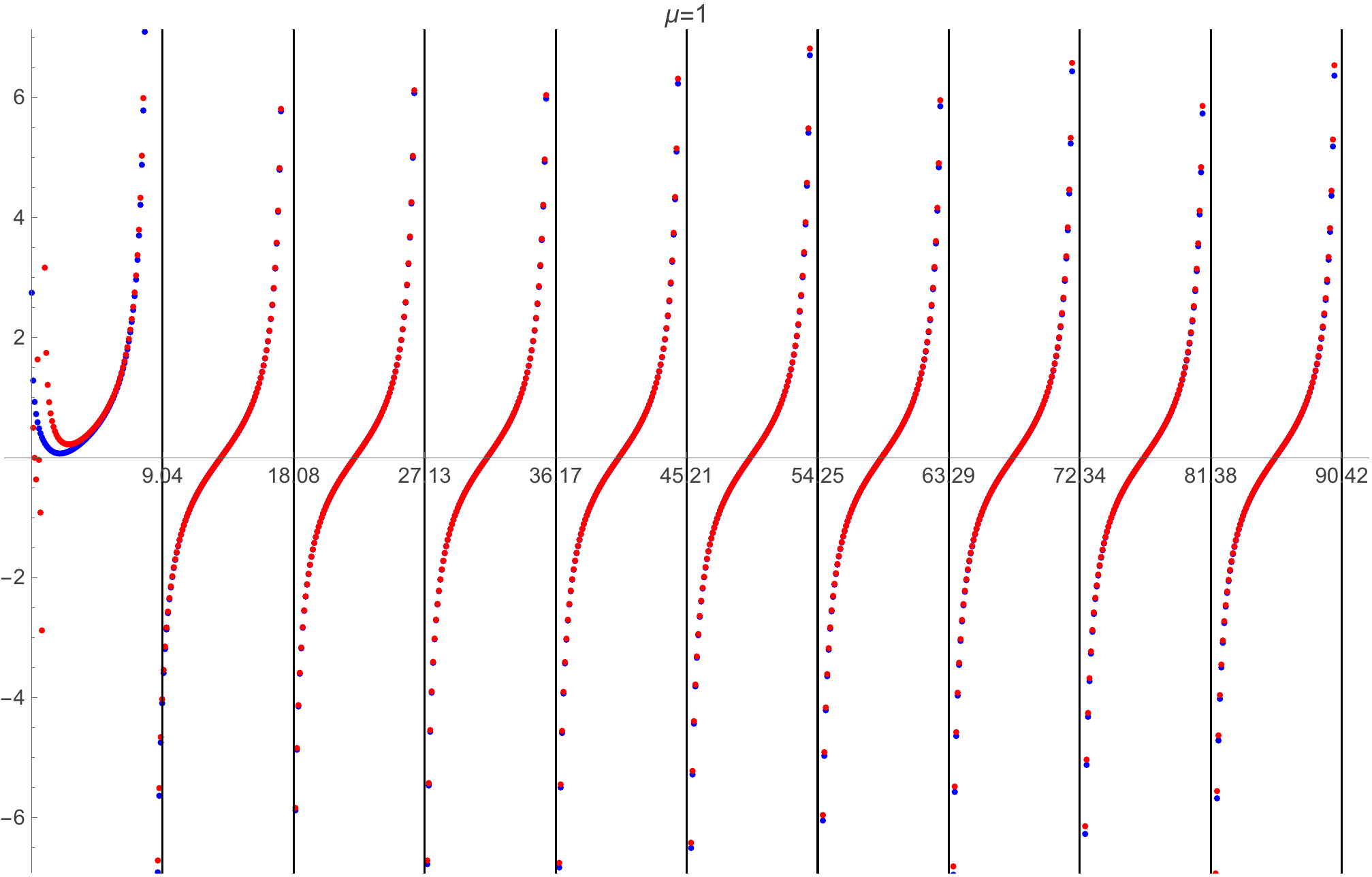}
\caption{The blue curve is the ratio $A_1/ A _2 $ obtained from the numerics, the red curve is $A _1 / A _2 $ obtained from (\ref{ratiopred}), (\ref{Thetaeq}). The black vertical lines show the predicted asymptotic spacing of the poles, $ \pi / I  (1)  $.}
\label{wkbratio}
\end{center}
\end{figure}

\begin{figure}[htbp]
\begin{center}
\includegraphics[scale=0.4]{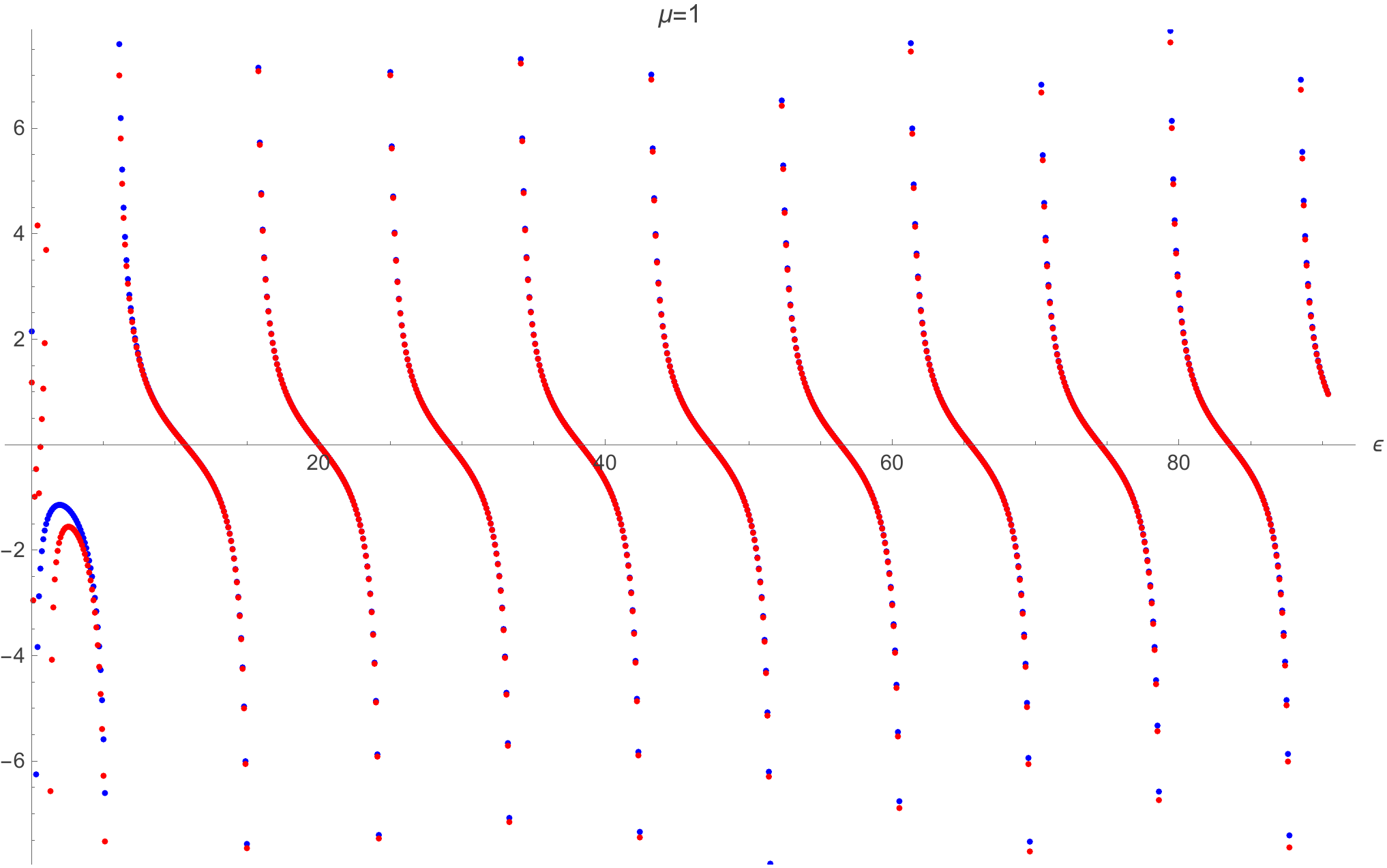}
\caption{The blue curve is the phase $\cot \Theta  $ obtained from the numeric solution of the ODE via (\ref{phaseshift}), the red curve is $\cot \Theta $ obtained from (\ref{Thetaeq}).}
\label{wkbphase}
\end{center}
\end{figure}

\newpage 
\subsection{The small-$\epsilon$ regime}
\label{small-eps}
Let us now consider the $\epsilon \ll 1 $ regime. We use a matched asymptotic expansion with two regions: near region $ 0 \leq x \lesssim 1$ with $x =O (1) $ at the right region boundary, far region $ \epsilon x \sim  1 $ with $x = O (1/ \epsilon )$ at the left boundary.

Note that the right boundary of the inner region has $x \sim 1 $ so we cannot simply take $\an $. Instead we  expand perturbatively in $\epsilon ^2 $.  Write 
\begin{equation}
a =a ^{(0)}  + \epsilon ^2 a ^{(2)} .
\end{equation}
At lowest order in $\epsilon $   (\ref{odek4bis}) becomes
\begin{equation} 
(a ^{(0)} ) ^{ \prime \prime } + P (a^{(0)} ) ^\prime =0
\end{equation} 
which can be integrated explicitly,
\begin{equation}
a^{(0)} (x) =C _2 \left( x ^2  + 4 \log x - \frac{1}{ x ^2 } \right) + C _1 .
\end{equation} 
Regularity at $x=0$ requires $C_2=0$, and we normalise $a^{(0)}(0)=1$, so $C _1 =1$. At  order $\epsilon ^2 $ we get the equation 
\begin{equation}
(a ^{(2)})^{ \prime \prime } + P (a^{(2)})  ^\prime  +  Q   =0.
\end{equation} 
This ODE can be integrated explicitly, but to do the matching we only need its behaviour for large $x$, which is
\begin{equation}
a ^{(2) }= -\frac{x ^2 }{2} \left(  \log x + \frac{\mu }{12} -1 \right) +O (\log ^2 x ) .
\end{equation} 
Hence in the near region, for large $x$, we have
\begin{equation}
a =1 -\frac{ \epsilon ^2 x ^2 }{2} \left(  \log x + \frac{\mu }{12} -1 \right)  + O (\epsilon ^4 ).
\end{equation} 

In the far region we have $x =O (1/ \epsilon ) \Rightarrow x \gg 1 $. Since for large $x$ we have $ Q \rightarrow 1 $ we can approximate $\epsilon ^2 Q $ with $\epsilon ^2 $, reobtaining (\ref{odefar}). To do the matching we need the small $x$ behaviour of $\af $, given by (\ref{afarnear}). The matching condition is then
\begin{equation}
1 -\frac{ \epsilon ^2 x ^2 }{2} \left(  \log x + \frac{\mu }{12} -1 \right)  
=- \frac{2 A _2 }{\pi \epsilon } +  \frac{ \epsilon  A _2}{ \pi }  x ^2 \log x+ \frac{\epsilon }{2 \pi } \left[ \pi A _1 + 2 A _2 \log (\epsilon /2 ) + ( 2 \gamma -1 ) A _2   \right] x ^2 .
\end{equation} 
Matching the constant term gives
\begin{equation}
A _2 =- \frac{\pi \epsilon }{2}
\end{equation} 
which also solves the $x ^2 \log x $ matching. Matching the $x ^2 $ term then gives $A _1 $, from which we get the ratio
\begin{equation}
\label{ratiosmallepsilon} 
\frac{A _1 }{A _2 }
=- \frac{2}{\pi } \log (\epsilon/2) + \frac{\mu }{6 \pi  } - \frac{(2 \gamma + 1)}{\pi }.
\end{equation} 
In particular, $ A _1 / A _2 $ diverges logarithmically as $\epsilon \rightarrow 0 ^+ $, with a slope which is independent of $\mu$. Plotting the result of the numerical integration of the full ODE shows full agreement with (\ref{ratiosmallepsilon}).
\begin{figure}[htbp]
\begin{center}
\includegraphics[scale=0.35]{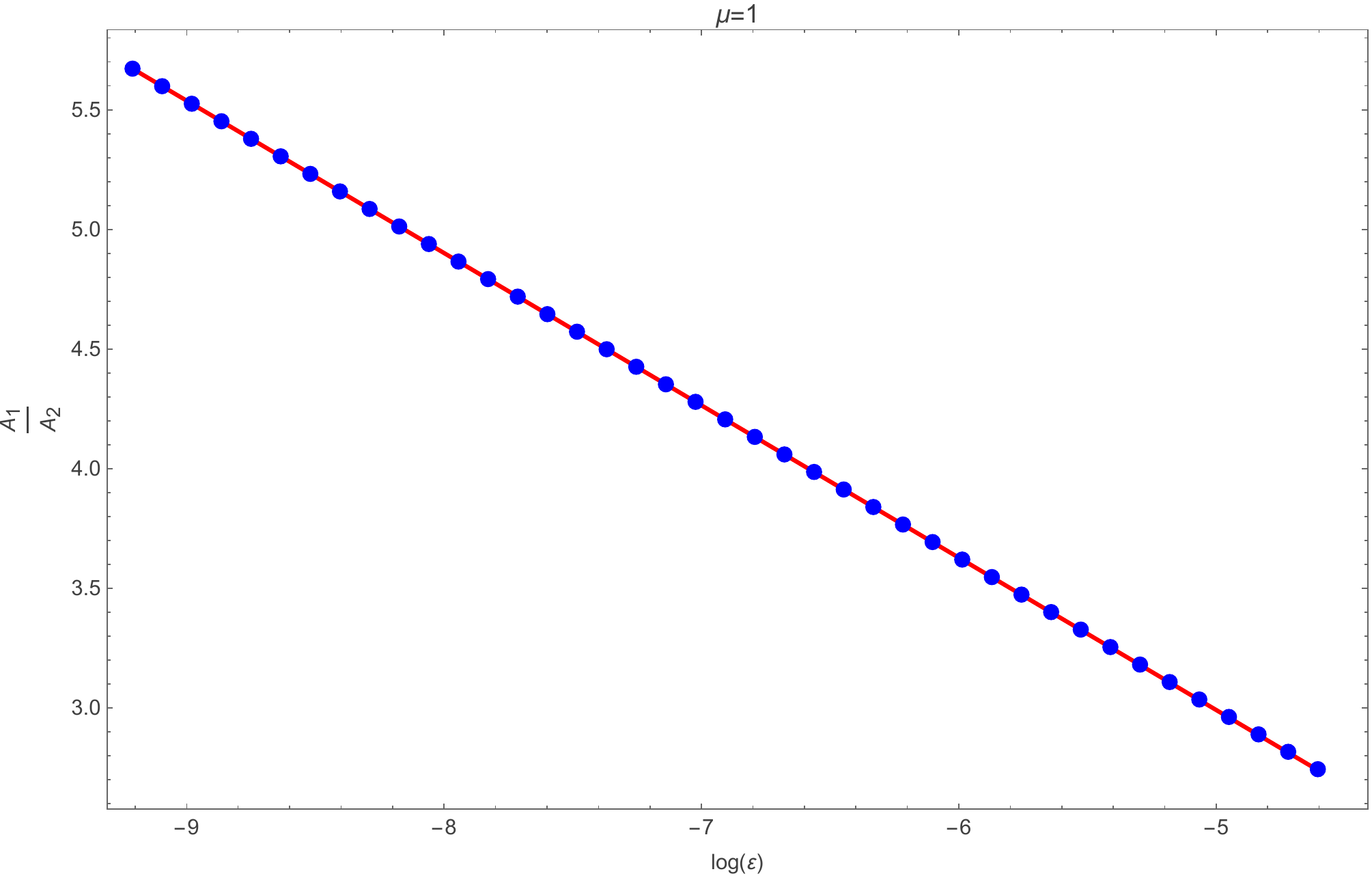}
\caption{The red curve  is the ratio $A_1/ A _2 $ obtained from (\ref{ratiosmallepsilon}) for $\mu =1 $, the blue dots show the ratio $A _1 / A _2 $ obtained numerically. Note that the values are plotted against $\log \epsilon$ for $\epsilon $ varying between $10 ^{ -4 } $ and $10 ^{ -2 } $.}
\end{center}
\end{figure}

\subsection{The low-energy effective action}

We can now write the low-energy effective action. This is the closest analogue to the zero-mode approximation dynamics \cite{Manton1982} that we can find when the zero mode is non-normalisable. We consider a  cutoff radius $\tilde{x}$ that is bigger than the size of the soliton and yet smaller than $1/\epsilon$
\begin{equation}
	1 \ll \tilde{x} \ll \frac{1}{\epsilon}
\end{equation}	 We take only the leading terms at low-energy, or large radius, from (\ref{effective1}) and we add a boundary term at $\tilde{x}$ obtaining
\begin{equation}
S_{\mathrm{eff}} = \int  \mathrm{d} \tau  \int_{\tilde{x}}^{\infty} \frac{ \mathrm{d} x }{x } \left(\dot{\alpha}(\tau ,x)^2 - \alpha^{ \prime }(\tau ,x)^2 \right) + \frac{M_{\mathrm{eff}}(\tilde{x})}{2} \int \mathrm{d} \tau \, \dot{\alpha}(\tau,\tilde{x} )^2 
\label{effective2}
\end{equation} 
%The analogy with a simpler known system is in Appendix \ref{simpleexample}. 
The boundary term is a localised kinetic term for $\alpha (\tau , \tilde{x})$. It is the only term which would be present if the zero mode were normalisable. 
%Here it is supplemented by a string $(1+1)$-dimensional action and has a scale running with the energy $\epsilon$.  
The boundary equation of motion from (\ref{effective2}) is  
\beq
 \frac{\alpha^{ \prime }(\tau,\tilde{x})}{\tilde{x}}  -  \frac{M_{\mathrm{eff}}(\tilde{x})}{2}  \ddot{\alpha}(\tau,\tilde{x})  = 0.
\label{condalpha}
\eeq
From (\ref{afarnear}) we get, to leading order,
\begin{equation} 
\frac{a ^\prime (\tilde{x})}{\tilde{x}}  = \frac{2 \epsilon A _2 }{\pi } \log \left(\frac{\tilde{x} \epsilon }{2} \right) + \epsilon  \left(A _1  +\frac{2 \gamma A _2 }{\pi }  \right),
\label{bceffective}
\end{equation}
and
 \beq
a(\tilde{x}) =   -\frac{2A_2}{\pi \epsilon },
 \eeq 
so we have a clear correspondence  between $A_{1,2}$ and $\alpha (\tau , x ) = a(x)  \mathrm{e} ^{i \epsilon \tau  }  $. 
 Substituting in (\ref{condalpha}) we obtain
 \beq
 \label{Meffrel} 
 \frac{2   A _2 }{\pi } \log \left(\frac{\tilde{x} \epsilon }{2} \right) +  A _1  +\frac{2 \gamma A _2 }{\pi }    -M_{\mathrm{eff}}(\tilde{x})\frac{A_2}{\pi } = 0.
 \eeq
 We can  match (\ref{Meffrel})  with (\ref{ratiosmallepsilon}) if we take the effective mass to be  
\beq
M_{\mathrm{eff}}(\tilde{x}) = 2 \log{\tilde{x}} +\frac{\mu}{6} -1.
\eeq
This value for $M_{\mathrm{eff}} $   is perfectly consistent with the effective mass of the zero mode computed from (\ref{actionalphafh}) and (\ref{ourf}) using $\tilde{x}$ as an IR cutoff:
\begin{align}
M_{\mathrm{eff}}(\tilde{x}) & = 2 \int_{0}^{\tilde x} 
\frac{x ^3  }{(1+ x^2)^2}\left( 1 +\frac{\mu }{(1 +x ^2 ) ^2 } \right) \mathrm{d} x \nn \\ 
& = \log(1+\tilde{x}^2) + \frac{\tilde{x}^2(\tilde{x}^4(\mu-6)+3\tilde{x}^2(\mu-4)-6)}{6 (1+\tilde{x}^2)^3}
= 2 \log \tilde x +\frac{\mu }{6}  -1 +O (\tilde x ^{-2} ) .
\end{align}

%It flows logarithmically to a Dirichlet boundary condition as $\epsilon \to 0 ^+$.  

\section{Conclusions and discussion}
\label{conclusion}
Non-normalisable zero modes fail to give a sensible $(0+1)$-dimensional effective action, but in the model considered here they give a $(1+1)$-dimensional effective action defined on a half-plane. At low energies these scattering zero modes can reach the soliton core, where the boundary conditions are determined. We have analyzed this scenario in the simplest possible model: the phase modulus of the holomorphic baby Skyrmion model. In future work we will consider more complex systems such as non-Abelian monopoles and monopoles in hyperbolic space. These systems are considerably more complex than the holomorphic baby Skyrmion.  For the non-abelian monopole it is known that,  in the Higgs phase,  the non-normalizable modes become  $1+1$ dynamical fields on the non-abelian magnetic vortex string \cite{Auzzi:2003em}.  This is somehow similar to what we discussed in this paper, although for non-abelian monopoles a version of the $1+1$ model in the Coulomb phase, like scattering of radial waves,  has never been found so far. 

There are similarities with the problem of fermionic fluctuations in a monopole background, known as the Callan–Rubakov effect \cite{Callan:1982ah,VARubakov_1988}, which can be treated, like the Kondo effect, as a boundary problem \cite{Affleck:1994aa}. In the presence of a magnetic monopole field, the angular momentum of fermions can be reduced to zero, producing an effective $s$-wave particle that can penetrate the monopole core. The anomaly is not suppressed in the monopole sector of the theory, giving rise to the non-conservation of fermion quantum numbers. Results concerning these phenomena have been obtained by reducing the theory to a model for an $s$-wave particle moving on a half-plane described by the radial and temporal coordinates. We have used a similar approach for non-normalisable zero modes in this work, focusing on scattering states of bosonic rather than fermionic fluctuations.

An interesting extension of this work would be to consider the limit in which the theory reduces to the pure nonlinear sigma model and the size modulus also becomes a non-normalisable zero mode. It would be interesting to see whether the $(1+1)$-dimensional effective approach can also be used for the size modulus, in particular to understand the nature of the finite-time singularities observed in lump dynamics. 

%It may be useful here to comment on some similarities, and difference. 

\subsection*{Acknowledgments}

%We thank K.~Konishi,  W.~J.~Zakrzewski for useful discussions. 
S.~B. and G.~F. are supported by the INFN special research project	grant ``GAST'' (Gauge and String Theories).  G.~F. acknowledges support from GNFM (Gruppo Nazionale per la Fisica Matematica).

\bibliographystyle{unsrt} 
\bibliography{holobaby-references}

\end{document}